\documentclass[pdflatex,sn-basic,iicol,Numbered]{sn-jnl}

\usepackage{multirow}%
\usepackage{amsmath,amssymb,amsfonts, xspace}%
\usepackage{amsthm}%
\usepackage{mathrsfs}%
\usepackage[title]{appendix}
\usepackage{textcomp}%
\usepackage{manyfoot}
\usepackage{algorithm}%
\usepackage{algorithmicx}%
\usepackage{algpseudocode}%
\usepackage{listings}%

\usepackage[moderate]{savetrees}
\usepackage{graphicx, epsfig, psfrag }
\usepackage{bm}
\usepackage{xcolor, xspace}
\usepackage{booktabs}
\usepackage{hyperref}
\usepackage{diagbox}
\usepackage{braket, xfrac}
\usepackage{soul, float, enumitem}
\usepackage[color=brown!60,textsize=scriptsize]{todonotes}
\def\cvs {CsV$_3$Sb$_5$\xspace}

\makeatletter
\renewcommand{\printhistory}{%
    \par\vspace{-30pt}%
    \begin{center}
    \fontsize{12bp}{9.5bp}\selectfont
    Dated: September 4, 2026
    \end{center}
    \par\vspace{-40pt}
}
\makeatletter
\long\def\@tablecaption#1#2{%
  \vbox{\hsize=\linewidth\noindent\tablecaptionfont{\bfseries #1}{\hskip2mm}#2\vphantom{y}\par}%
}
\makeatother

\begin{document}
  \title[Article Title]{Comment on: \textit{Microscopic signatures of an imaginary charge density wave in a kagome metal}}
 
\author[1]{\fnm{Ilija K.} \sur{Nikolov}}

\author[1, 2]{\fnm{Adrien} \sur{Rosuel}}

\author*[1, 2]{\fnm{Vesna F.} \sur{Mitrovi\'c}}\email{vemi@brown.edu}

\affil[1]{\orgdiv{Department of Physics}, \orgname{Brown University}, \orgaddress{\city{Providence}, \postcode{02912}, \state{Rhode Island}, \country{USA}}}

\affil[2]{\orgdiv{Brown Center for Theoretical Physics and Innovation}, \orgname{ Brown University}, \orgaddress{\city{Providence}, \postcode{02912}, \state{Rhode Island}, \country{USA}}\vspace{-10pt}}

\maketitle

\section{Introduction}

In Ref.~\cite{Suetsugu2026}, Suetsugu \textit{et al.} presented nuclear quadrupole resonance (NQR) and Zeeman-perturbed NQR (Zp-NQR) measurements on the kagome superconductor CsV$_3$Sb$_5$, reporting microscopic signatures of an imaginary charge density wave (iCDW) and time-reversal symmetry breaking (TRSB) emerging at a remarkably high temperature $T^* \approx 120$~K. This conclusion is based on two observations: an apparent enhancement of the NQR linewidth at~$T^*$ and asymmetric Zp-NQR lineshapes attributed to emergent hyperfine fields from loop currents. Notably, the observation of a linewidth enhancement depends on the expectation of a nematic transition at $T^*$~\cite{Asaba2024}, which remains contested~\cite{Frachet2024, Liu2024, Guo2024}.

In this Comment, we demonstrate through exact diagonalization and quantitative analysis that the reported signatures of loop currents in Ref.~\cite{Suetsugu2026} can instead be fully explained by crystalline mosaicity and pre-transitional CDW fluctuations. Specifically: (i) the NQR linewidth tracks a standard Curie-Weiss (CW) law with no distinct anomaly at $T^*$; (ii) the asymmetric Zp-NQR lineshapes stem from an angular mosaic spread, naturally reproducing the observed spin-transition-dependent skewness; and (iii) the reported emergent field ($h_{\text{loc}}$) arises from undisclosed fitting constraints based on physically ungrounded assumptions, and the misattribution of angular distributions to internal fields.  

Here, we address the Reply by Suetsugu \textit{et al.}~\cite{firstreply}, showing that it fails to resolve any of our critical objections. We demonstrate that the arguments in Ref.~\cite{firstreply} rely on an idealized definition of mosaicity that ignores strain-mediated lattice distortions that are well-documented in CDW systems via NQR, STM, and x-ray scattering~\cite{CBerthier1978, Ghoshray2009, Arguello2014, Wu2015, Vinograd2019, Liu2021, Chen2022, Subires2023, Feng2023, Nikolov2026}. Most critically, we confirm that when the full nuclear-spin Hamiltonian is exactly diagonalized without approximations and the unjustified constraints are removed, the inferred loop-current signature vanishes ($h_{\text{loc}} \to 0$). We therefore conclude that the claim of an iCDW state is unsupported by the experimental data of Ref.~\cite{Suetsugu2026} and is instead a consequence of incomplete spectral simulation and inadequate treatment of the standard Zp-NQR physics.

\section{Absence of anomaly in NQR linewidth}

\begin{figure}[t]  
\includegraphics[width=\linewidth]{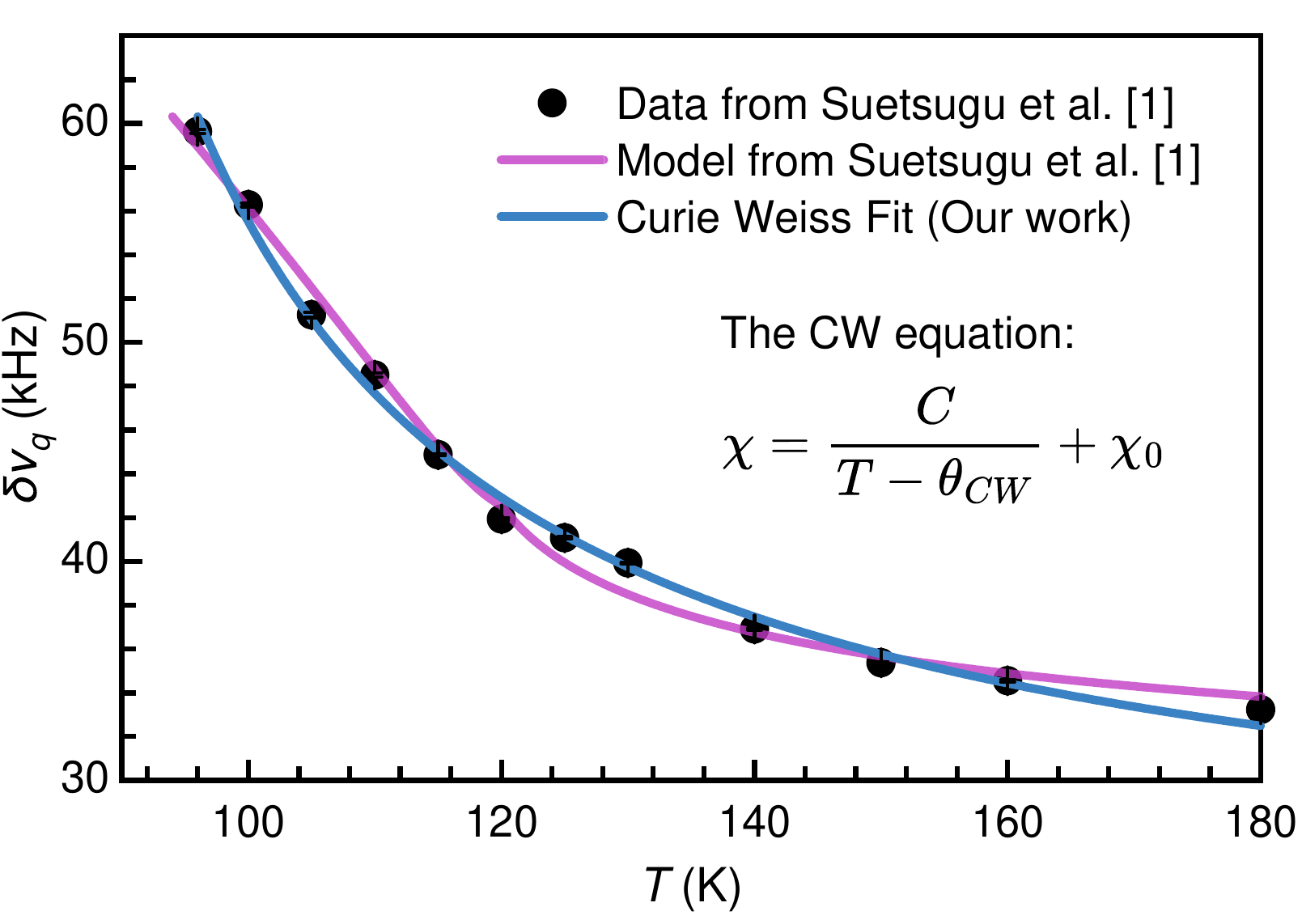}
 \vspace{-0.5cm}
\caption{ \textbf{Absence of a linewidth anomaly at 120~K.} The $^{121}$Sb2 linewidth continuously tracks a standard Curie-Weiss law (blue line, $\theta_\text{CW} \sim 70$~K) driven by incipient CDW fluctuations~\cite{Feng2023, Nikolov2026}, with no distinct enhancement at~$T^*~\sim~120$~K. The thermal fluctuation model from Ref.~\cite{Suetsugu2026} deviates from the data most noticeably within the interval~$120$~K~$\lesssim~T~\lesssim$~140~K.}
\label{fig:fig_CW}
\vspace{-0.4cm}
\end{figure}

\begin{figure*}[t] 
\includegraphics[width=\linewidth]{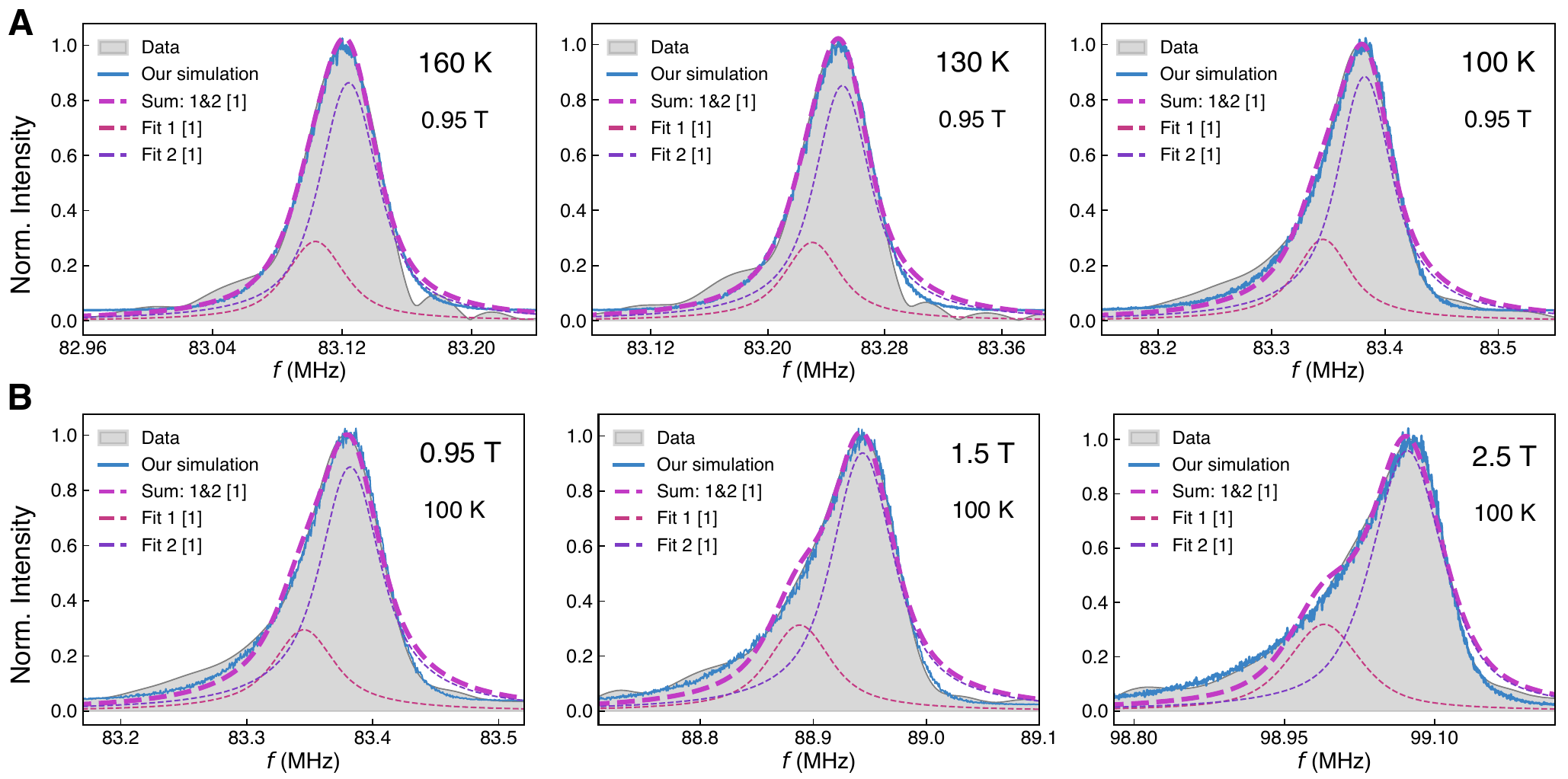}
\caption{ \textbf{Mosaicity reproduces the asymmetric lineshapes.} 
Experimental spectra (shaded gray area) for the $^{121}$Sb2 site in the  $-1/2 \leftrightarrow -3/2$ transition are compared with our mosaicity-only simulation (blue line) and the loop-current model from Ref.~\cite{Suetsugu2026} (colored dashed lines). The mosaicity model, $\delta\theta~(160~{\rm K}) \approx 1.1^\circ\text{ -- }\delta\theta~(100~{\rm K})\approx1.6^\circ$ (no field dependence), captures the data well without emergent fields. The loop-current model predicts distinct peak structures absent in the data at higher fields (1.5~T \& 2.5~T). The increase in $\delta\theta$ upon cooling beyond the baseline mosaicity of $1.1^\circ$ signals the additional orientational distribution of the local EFG principal axes caused by the nucleation of CDW puddles, hallmark of the incipient CDW~\cite{Chen2022, Subires2023, Feng2023, Nikolov2026}.}
\label{fig:Fig_sim_fit}
\end{figure*}

In Fig.~\ref{fig:fig_CW}, we re-examine the temperature-dependent $^{121}$Sb2 NQR linewidth from Ref.~\cite{Suetsugu2026}. Although the authors reported a broadening enhancement near~120~K, the established literature on CsV$_3$Sb$_5$ has consistently observed continuous pre-transitional broadening, even under pressure and doping~\cite{Feng2023, Nikolov2026}. As demonstrated by the standard CW fit in Fig.~\ref{fig:fig_CW}, we find no clear anomaly near $T^*~\sim~120$~K that would signal a phase transition. Rigorously establishing such a feature necessitates evaluating the first derivative, which is precluded by the temperature step-size used~\cite{Suetsugu2026}. Furthermore, the authors argued that CDW fluctuations are not strong in a first-order transition~\cite{Suetsugu2026}. This view is inconsistent with the existing literature showing pre-transitional CDW fluctuations in \cvs, well above $T^*~\sim~120$~K, regardless of the transition's thermodynamic order~\cite{Chen2022, Subires2023, Feng2023, Nikolov2026}. Therefore, the available data from Ref.~\cite{Suetsugu2026} do not provide sufficient evidence for a distinct anomaly at 120~K.

\section{Asymmetric lineshapes in Zp-NQR}
The frequencies $\nu_Q \approx 75$~MHz vs. $\nu_L \approx 5$~MHz at 0.5~T place the system strictly in the Zp-NQR regime. In this limit, the subdominant Zeeman interaction mixes the quadrupolar eigenstates in an orientation-dependent manner, so the full Hamiltonian is required when the field is not aligned with a principal electric field gradient~(EFG) axis (Appendix~\ref{app:hamil}). A perfect alignment across a macroscopic crystal is not experimentally attainable because of mosaicity (Appendix~\ref{app:crystal_qual}). Furthermore, as CsV$_3$Sb$_5$ approaches the CDW transition temperature~($T_{\rm CDW}$), the nucleation of short-range CDW puddles naturally enhances spatial variations in both EFG magnitude and orientation~\cite{Chen2022, Subires2023, Feng2023, Nikolov2026}. We show that without accounting for structural mosaicity, the resulting asymmetric lineshapes are inadvertently attributed to emergent hyperfine fields from loop currents~\cite{Suetsugu2026}.

Mathematically, the Hamiltonian in Ref.~\cite{Suetsugu2026} rotates the Zeeman term but leaves the detection operator~($I_+$) unrotated, which implicitly treats the laboratory RF coil orientation as co-rotating with the local EFG axis of each crystallite.
We correct this by performing exact diagonalization of the full Hamiltonian~(Eq.~\ref{eq:H}) with the Zeeman and RF fields fixed in the laboratory frame, while rotating the local EFG axes to model mosaicity  (code available online\footnote{\url{https://github.com/ilijanikolov/zpnqr_simulation}}). We show the specific impact on the lineshape of each Hamiltonian term in Extended Data Fig.~\ref{fig_supp:Fig_sim}.

We demonstrate that the observed spectral skewness can arise from an angular distribution of EFG axes. Crucially, this geometric effect is transition-dependent, producing the opposite left/right skewness for the $-1/2 \leftrightarrow -3/2$ and $+1/2 \leftrightarrow +3/2$ transitions observed experimentally. Incorporating realistic distribution widths ($\delta\theta \approx 1.1^\circ\text{--}1.6^\circ$) reproduces the data with high fidelity across all fields and temperatures (Fig.~\ref{fig:Fig_sim_fit}). Importantly, we use $\delta\theta$ to denote the distribution of local EFG principal-axis orientations. At high temperature, this distribution may be dominated by conventional crystallographic mosaicity, while additional local structural distortions associated with CDW correlations can contribute on approaching $T_\text{CDW}$~\cite{Feng2023, Subires2023, Nikolov2026}. Thus, the asymmetric lineshapes are quantitatively reproduced by conventional Zp-NQR physics in a realistic crystal with baseline mosaicity enhanced by pre-transitional CDW puddles, without invoking TRSB and emergent hyperfine fields.

\section{Influence of fitting constraints on extracted $h_{\rm loc}$}
As demonstrated in Fig.~\ref{fig:Fig_sim_fit} and Extended Data Fig.~\ref{fig:supp_sb1}, the triple-\textbf{q} loop-current model proposed in Ref.~\cite{Suetsugu2026} systematically deviates from the observed spectra. The experimental lineshapes remain continuous and broad, lacking any features indicative of unresolved splitting from emergent hyperfine fields. This discrepancy is most pronounced at~2.5~T and~100~K, where the distinct peak structures predicted by the loop-current model (dashed red lines in Fig.~\ref{fig:Fig_sim_fit}) are not resolved in the experimental spectra.

At zero external field, a rigid rotation of the EFG tensor does not change the NQR transition frequencies because there is no external field direction with which to form an angular dependence. The frequency broadening is therefore governed primarily by variations in the EFG parameters. Consequently, zero-field broadening arises from variations in EFG magnitude and asymmetry (see simulations online), resulting in the symmetric lineshapes observed experimentally~\cite{Feng2023, Nikolov2026, Suetsugu2026}.

\begin{figure}[t] 
\vspace{-.5cm}
\includegraphics[width=\linewidth]{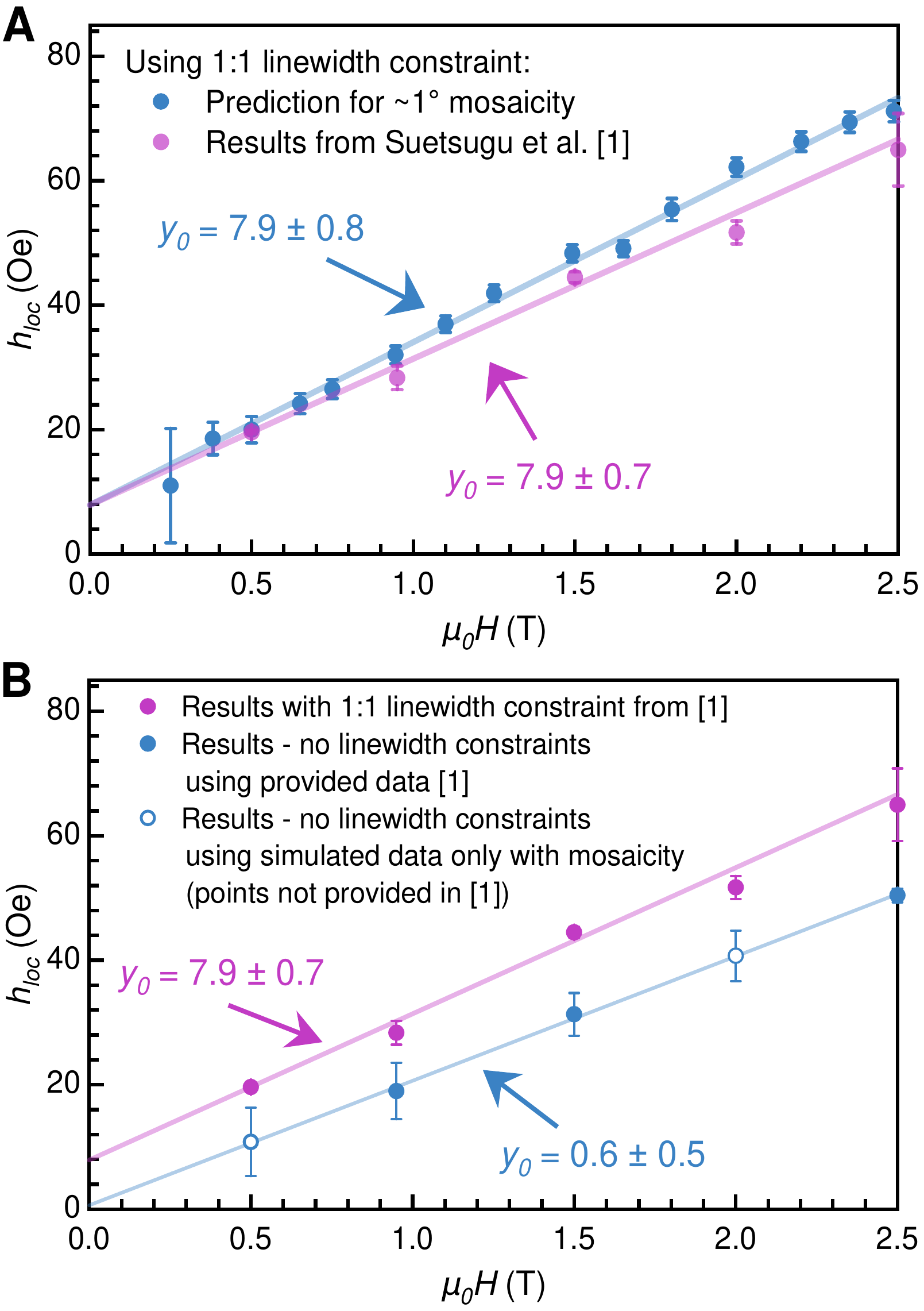}
\vspace{-.4cm}
\caption{\textbf{Origin of the apparent zero-field $h_{loc}$ intercept.} 
\textbf{A} Applying the constrained fitting procedure from Ref.~\cite{Suetsugu2026} to a mosaicity-only simulation yields a finite intercept ($h_{\rm loc} \approx 7.9$~Oe), revealing that structural mosaicity alone can mimic an apparent local field.
\textbf{B} When the restrictive equal-broadening constraint is relaxed, the extracted $h_{\rm loc}$ extrapolates to zero within error, consistent with conventional Zeeman scaling in Zp-NQR and the absence of spontaneous TRSB.}
\vspace{-0.3cm}
\label{fig:Fig_field}
\end{figure}

In Fig.~\ref{fig:Fig_field}, we demonstrate that the apparent zero-field intercept of $h_{\rm loc}$~\cite{Suetsugu2026} arises from both the restrictive fitting constraints and the implicit assumption that all local EFG axes remain perfectly aligned with the crystallographic $c$-axis. First, applying the fitting procedure described in Ref.~\cite{Suetsugu2026} to our mosaicity-only simulations reproduces the reported intercept within uncertainty, $h_{\rm loc}^{\rm mosaic} \approx 7.9$~Oe (Fig.~\ref{fig:Fig_field}A). Strikingly, the fitting procedure in~\cite{Suetsugu2026} is constrained by an unreported equal-linewidth assumption (1:1 ratio) between the two magnetically inequivalent sites (Appendix~\ref{app:hyp_field}). As shown in Fig.~\ref{fig:Fig_field}B, relaxing the restrictive broadening constraint while enforcing the physically realistic~$3\!:\!1$ intensity ratio causes the extracted~$h_{\rm loc}$ to extrapolate to 0~Oe within error. However, applying the constraint~of~$3\!:\!1$ for both intensity and width leads the fitting to converge on a resolved spectral splitting absent in the experimental data (Extended Data Fig.~\ref{fig:supp_constraints}). Thus, the reported finite $h_{\rm loc}$ is highly sensitive to the fitting constraints, preventing its unique attribution to emergent loop-current fields.

\section{Reply to \href{https://arxiv.org/abs/2608.24927v1}{arXiv:2608.24927v1}}
In their Reply~\cite{firstreply}, Suetsugu \textit{et al.} focus narrowly on the definition of mosaicity and the distribution of the EFG asymmetry parameter $\eta$. However, this narrow focus leaves three critical issues unaddressed, each directly undermining the proposed iCDW scenario: 
\begin{enumerate}[label=\roman*)]
    \item \textbf{Absence of a phase transition:} The temperature dependence of the local susceptibility follows a standard CW law and lacks the anomalies required for a phase transition at $T^*$;
    \item \textbf{Unjustified fitting constraints:} The non-zero local field $h_{\text{loc}}$ in the limit of $H \rightarrow 0$ is an artifact of physically ungrounded assumptions and fitting constraints not explicitly disclosed in Refs.~\cite{Suetsugu2026, firstreply}; and
    \item \textbf{Inadequate theoretical modeling:} The use of approximate models rather than the exact diagonalization of the full nuclear spin Hamiltonian.  
\end{enumerate}

Suetsugu~\textit{et al.}'s argument regarding $\eta$~\cite{firstreply} rests on an idealized definition of mosaicity as a rigid rotation of a perfect lattice and does not account for the well-established coupling between lattice strain and local electronic environments in real crystals. Consequently, Suetsugu~\textit{et al.}'s dismissal of temperature-dependent orientation distributions conflates intrinsic CDW domain percolation with static structural defects, at odds with established literature on \cvs~\cite{Chen2022, Feng2023, Subires2023, Nikolov2026}. More critically, the Reply~\cite{firstreply} does not address the core methodological limitation of using an incompletely diagonalized Hamiltonian and thus fails to invalidate the conventional explanation for the observed spectra in Ref.~\cite{Suetsugu2026}. In this section, we demonstrate that the objections of Ref.~\cite{firstreply} are readily resolved within the standard physical framework without invoking an iCDW state to account for the spectroscopic observations in Ref.~\cite{Suetsugu2026}. Thus, the arguments presented in the Reply~\cite{firstreply} do not establish the existence of an iCDW state, nor do they rule out the standard Zp-NQR physical origin of the spectroscopic observations. 

First, the expectation that parameter distributions must be identical across distinct crystallographic sites assumes a level of homogeneity rarely found in systems proximate to structural instabilities. In such materials, local strain fields couple differently to distinct sites based on their specific symmetry and bonding environments~\cite{Frassineti2023}. What is described as an inconsistency in our model is, in fact, a physical signature of strain-mediated mosaicity, where local lattice distortions naturally induce site-dependent variations in both orientation ($\delta\theta$) and the EFG tensor ($\delta\eta$). Nevertheless, even if one imposes identical distributions for both $\delta\eta$ and $\delta\theta$ across the Sb1 and Sb2 sites at~100~K, the Sb1 lineshape is accurately reproduced by allowing variations in the quadrupolar frequency magnitude distribution,~$\delta\omega_Q$ (Extended Data Fig.~\ref{fig:supp_sb1_same_params}). This demonstrates that the spectral features are robustly explained by standard physical parameter distributions, rendering the site-selective objection raised in Ref.~\cite{firstreply} non-essential.

Furthermore, the observed temperature dependence of $\delta\omega_Q$ and $\delta\theta$ is not an anomaly but an expected consequence of structural disorder driven by quasi-static, pre-transitional CDW fluctuations upon cooling toward~$T_{\text{CDW}}$~\cite{firstreply}. As the system approaches the transition, the nucleation of CDW domains induces local lattice distortions that modulate the effective mosaic spread and EFG parameters. These mechanisms are extensively documented across CDW systems, including \cvs, by NQR/NMR~\cite{CBerthier1978, Ghoshray2009, Wu2015, Vinograd2019, Feng2023, Nikolov2026} and STM~\cite{Arguello2014, Liu2021}. Distinctly for \cvs, x-ray scattering confirms an order-disorder transition with the critical growth of quasi-static CDW domains above $T_{\text{CDW}}$~\cite{Chen2022, Subires2023}, providing direct experimental support for the physical basis of our temperature-dependent parameters.

The decisive factor in this debate, however, lies in the theoretical treatment of the Hamiltonian and the fitting procedure.
The signatures reported by Suetsugu \textit{et al.}~\cite{Suetsugu2026} are mathematical artifacts arising directly from the use of a perturbative and incompletely diagonalized Hamiltonian. When the full Hamiltonian is solved exactly, these artifacts vanish.
Specifically,  when the \textbf{full NMR Hamiltonian is exactly diagonalized}, with all quadrupole and magnetic interactions and their physically realistic distributions included without perturbative truncation, the apparent spectral asymmetries and site-selective anomalies attributed to an iCDW vanish. Moreover, when the fitting is performed without imposing the unphysical constraints that cause the fit to generate apparent spectral features absent from the experimental data (Fig.~\ref{fig:Fig_sim_fit} and Extended Data Fig.~\ref{fig:supp_constraints}), the finite local field scenario likewise disappears. Notably, these constraints and their role were not explicitly reported in Refs.~\cite{Suetsugu2026, firstreply} and can only be uncovered by reverse engineering the fitting procedure; see the GitHub repository for detailed analysis\footnote{\url{https://github.com/ilijanikolov/zpnqr_simulation}}. We thus conclude that the features interpreted as exotic physics are mathematical artifacts of an approximate solution. Until it can be demonstrated that these results persist under exact diagonalization with an appropriate realistic model, the claim of an iCDW state is contradicted by a rigorous re-analysis, which demonstrates that the reported signatures are merely consequences of incomplete spectral simulation. 

\section{Conclusion}

In this Comment, we demonstrate that the spectra reported in Ref.~\cite{Suetsugu2026} are fully consistent with a conventional physical framework, rendering the invocation of an iCDW phase unnecessary. We show that the experimental lineshapes arise naturally from standard Zp-NQR effects in a realistic crystal exhibiting mosaicity and strain-induced EFG distributions. The linewidth evolution follows the CW law characteristic of incipient CDW fluctuations, a well-established precursor phenomenon~\cite{CBerthier1978, Ghoshray2009, Arguello2014, Wu2015, Vinograd2019, Liu2021, Chen2022, Subires2023, Feng2023, Nikolov2026}.

Most critically, we find that the reported finite internal field $h_{\text{loc}}$ at zero external field, the hallmark of TRSB, is a mathematical artifact resulting from the restrictive fitting constraints and the use of an approximate Hamiltonian. These fitting constraints, which were not explicitly reported in Refs.~\cite{Suetsugu2026, firstreply}, generate an apparent finite $h_{\text{loc}}$ that is absent from the experimental data. When these unphysical constraints are removed and the \textbf{full NMR Hamiltonian is exactly diagonalized}, the inferred loop-current signature vanishes, with $h_{\text{loc}} \to 0$ as $H \to 0$. This shows that the reported signatures of spontaneous TRSB do not require a phase transition into a triple-\textbf{q} loop-current state, but instead arises from inadequate spectral modeling and constrained fitting. We therefore conclude that the data reported in Ref.~\cite{Suetsugu2026} do not require an unconventional iCDW interpretation. Rather, the observations are fully described by the standard physics of Zp-NQR in a lattice with realistic mosaicity, quenched disorder, and an incipient CDW instability.

\section*{Data and code availability}
\vspace{-.2cm}
The code and simulations are available on GitHub: \url{https://github.com/ilijanikolov/zpnqr_simulation}.
The experimental data are from~\cite{Suetsugu2026} and are not hosted on the GitHub repository.

\section*{Acknowledgments}
\vspace{-.2cm}
We thank Aaron Hui for helpful discussions.

\section*{Competing interests}
\vspace{-.2cm}
The authors declare no competing interests.

\section*{Author contributions}
\vspace{-.2cm}
I.K.N. wrote the code, analyzed the data, and performed the simulations and fitting. I.K.N., A.R. and V.F.M. interpreted the data. V.F.M. provided conceptual advice. All authors
discussed the simulation and analysis, commented on and edited the manuscript.

\begin{appendices}
\section*{Extended data and appendices}
\renewcommand{\figurename}{Extended Data Fig.}
\section{Spin Hamiltonian}
\label{app:hamil}
\begin{figure*}[t] 
\includegraphics[width=\linewidth]{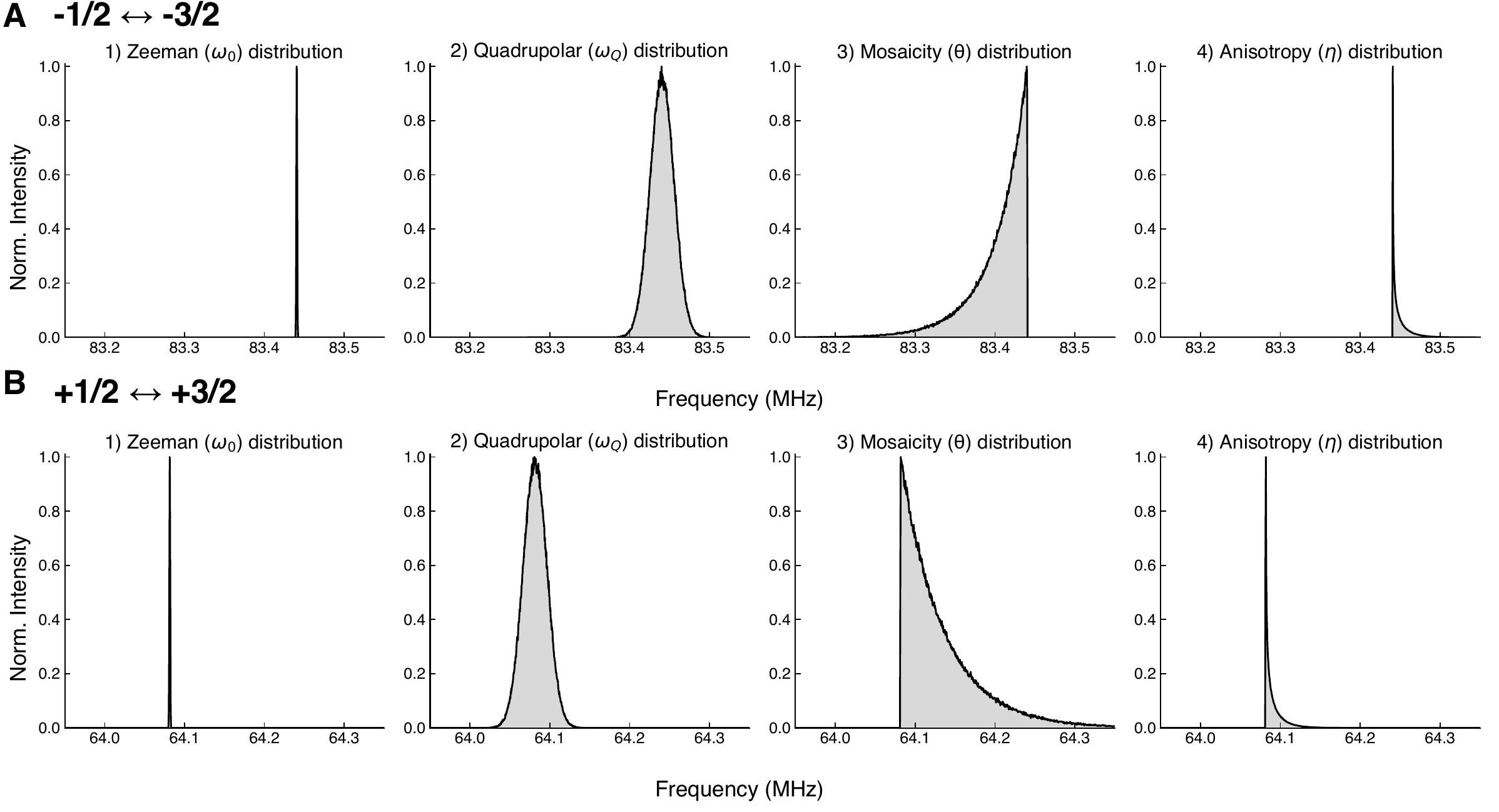}
\vspace{-.5cm}
\caption{ \textbf{Mosaicity and spectral asymmetry.} 
Simulating the linewidth using a static distribution of the different parameters of the Hamiltonian~\ref{eq:H}. Only the angular distribution of the EFG principal axis ($\theta$, representing mosaicity) reproduces the opposite skewness observed between the $-1/2 \leftrightarrow -3/2$ and $+1/2 \leftrightarrow +3/2$ transitions. A distribution in the magnitude of the Zeeman $\omega_0$ and quadrupolar frequency $\omega_Q$ yields symmetric broadening, while a distribution in magnitude of the asymmetry parameter $\eta$ yields only right skewness. The parameters used, $\delta\theta \approx 1.6^\circ$, $\delta\omega_Q \approx 5$~kHz, reflect the combined effects of intrinsic mosaicity and incipient CDW puddles at 100~K~\cite{Subires2023, Nikolov2026}. Field inhomogeneity ($\sim 50~\mu$T) is negligible.}
\label{fig_supp:Fig_sim}
\vspace{-0.3cm}
\end{figure*}
Following Ref.~\cite{Vega2010}, the Hamiltonian of the Zeeman-perturbed NQR can be written as
\begin{equation}
    		\label{eq:H}
    		\mathcal{H} = H_Z + H_Q,
	\end{equation}
where the orientation of the field is taken as the $z$-axis
\begin{equation}
     H_Z = \omega_0 I_z,
\end{equation}
whereas the quadrupolar interaction in the Cartesian coordinates is
\begin{equation}
\begin{split}
\label{eq:HQ}
\hat{\mathcal{H}}_{\text{Q}} = & \frac{eQ}{4I(2I-1)\hbar} \{ V_{zz}[3\hat{I}_{z}^{2} - I(I + 1)] \\
& + (V_{xx} - V_{yy})(\hat{I}_{x}^{2} - \hat{I}_{y}^{2}) + 2V_{xy}(\hat{I}_{x}\hat{I}_{y} + \hat{I}_{y}\hat{I}_{x}) \\
& + 2V_{xz}(\hat{I}_{x}\hat{I}_{z} + \hat{I}_{z}\hat{I}_{x}) + 2V_{yz}(\hat{I}_{y}\hat{I}_{z} + \hat{I}_{z}\hat{I}_{y}) \}
\end{split}
\end{equation}
where
\begin{equation}
\begin{split}
V_{xx} &= \frac{1}{2}\;eq\left(3\;\sin^2\theta - 1 - \eta\;\cos^2\theta\;\cos 2\phi\right), \\[0.5ex]
V_{yy} &= \frac{1}{2}\;eq\left(-1 + \eta\;\cos 2\phi\right), \\[0.5ex]
V_{zz} &= \frac{1}{2}\;eq\left(3\;\cos^2\theta - 1 - \eta\;\sin^2\theta\;\cos 2\phi\right), \\[0.5ex]
V_{xy} &= V_{yx}
= \frac{1}{2}\;eq\;\eta\;\cos\theta\;\sin 2\phi, \\[0.5ex]
V_{xz} &= V_{zx}
= -\frac{1}{2}\;eq\;\sin\theta\;\cos\theta\left(3 + \eta\;\cos 2\phi\right), \\[0.5ex]
V_{yz} &= V_{zy}
= \frac{1}{2}\;eq\;\eta\;\sin\theta\;\sin 2\phi,
\end{split}
\end{equation}
and $\theta, \phi$  are the polar and azimuthal angles with respect to the applied field. The asymmetry parameter is defined as 
\begin{equation}
    \eta = \frac{V_{yy} - V_{xx}}{V_{zz}},
\end{equation}
following the convention $|V_{zz}| \ge |V_{xx}| \ge |V_{yy}|$.
For convenience, we define the parameter:
\begin{equation}
    \omega_Q = \frac{eQ}{8I(2I-1)\hbar}.
\end{equation}
In our simulation, the FWHM is $\delta \eta \sim 0.01 $, which means that the effect of $\phi$ is subdominant, confirmed by calculations. For simplicity, we keep $\phi$ at zero, as having a uniform distribution of $\phi$ from 0 to 2$\pi$ only slightly changes the fitting parameters, mostly the spectrometer offset, but does not affect the lineshapes (see simulations online for details). 

In general, if one also included the orientational distribution of the applied field or the local field induced by the hyperfine interaction, one would need to modify the $H_Z$ term:
\begin{equation}
\begin{split}
        H_Z(\theta, \phi) = &- \gamma \hbar \mu_0(H + \Delta H) \\
        &\times \left(I_z\cos\theta + \frac{I_+ e^{-i\phi} + I_- e^{i\phi}}{2}\sin\theta\right).
\end{split}
\end{equation}
Nowadays, applied magnetic fields are very homogeneous so that the Zeeman distribution is much weaker than the mosaicity-induced one. In addition, we show that we do not need to invoke local emergent fields to explain the data.

\begin{figure}[t] 
\includegraphics[width=\linewidth]{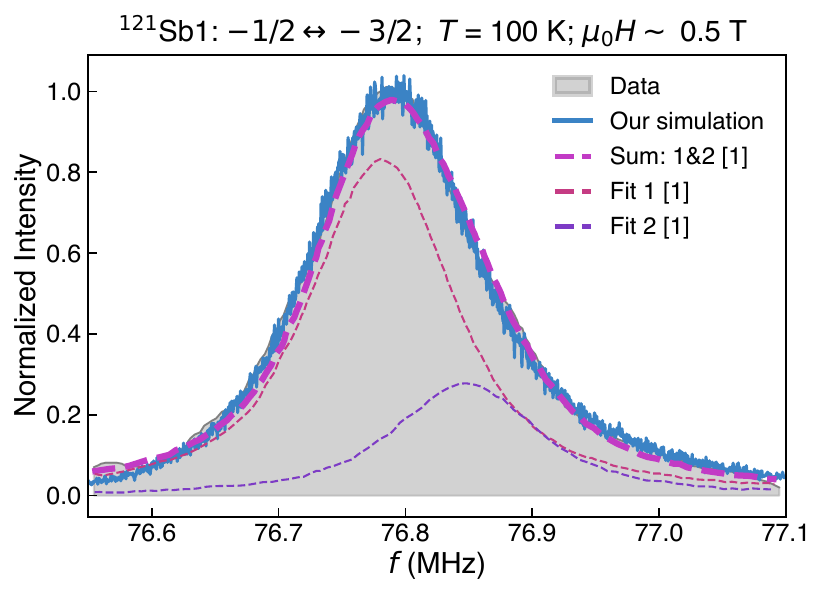}
\vspace{-.5cm}
\caption{\textbf{Enhanced disorder at the Sb1 site.} 
The Sb1 site exhibits significantly broader spectra than Sb2, consistent with greater structural disorder~\cite{Frassineti2023, Feng2023, Nikolov2026, Suetsugu2026}. Our simulations reproduce this behavior using larger distribution widths for Sb1 $\delta\omega_Q \approx 13.5$~kHz, $\delta\theta \approx 2.8^\circ$, $\delta\eta \approx 0.029$ compared to Sb2, confirming that the Sb1 environment is more sensitive to local lattice distortions and CDW formation.}
\label{fig:supp_sb1}
\vspace{-0.2cm}
\end{figure}

\begin{figure}[t] 
\includegraphics[width=\linewidth]{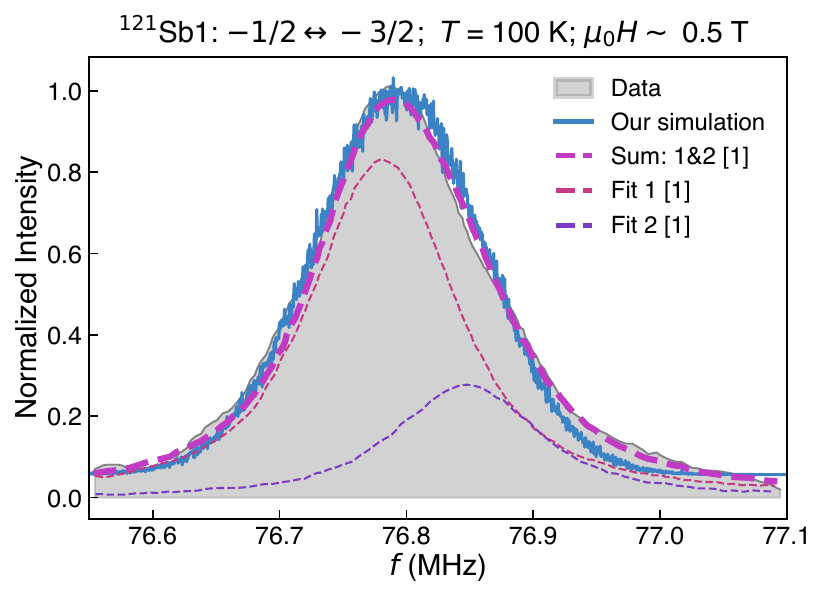}
\vspace{-.5cm}
\caption{\textbf{Furthrer confirmation for structural disorder at Sb1 site.} 
Our simulations also reproduce the 100~K Sb1 lineshapes by setting $\delta\omega_Q \approx 21$~kHz while keeping other parameters identical to those of the Sb2 site at 100~K, $\delta\theta \approx 1.6^\circ$ and $\delta\eta \approx 0.01$. Note that the baseline offset is larger for Sb1 when compared to Sb2. }
\label{fig:supp_sb1_same_params}
\vspace{-0.2cm}
\end{figure}

\section{Crystal quality}
\label{app:crystal_qual}
In the pristine phase, the structural NQR intensity ratio between Sb1 (in-plane) and Sb2 (out-of-plane) sites is fixed at~$1\!:\!4$, established by theory and experiment~\cite{Frassineti2023, Wang2023, Nikolov2026}. The ratio of~$1\!:\!1.3$ reported in Ref.~\cite{Suetsugu2026} deviates substantially from the expected ratio of~$1\!:\!4$. While the authors attribute this discrepancy to frequency-dependent sensitivity and varying excitation conditions, standard corrections appear insufficient. First, normalizing the spectral intensity by $\nu^2$ to account for the Boltzmann distribution and Faraday induction yields only a minimal adjustment between the two frequencies, at most:
\begin{equation}
    \left(\frac{\nu_{^{121}\text{Sb2, NQR1}}}{\nu_{^{121}\text{Sb1, NQR1}}}\right)^2  \approx\left(\frac{73.75 \text{ MHz}}{71.75 \text{ MHz}}\right)^2 \approx 1.057.
\end{equation}
Furthermore, variations in excitation conditions are typically mitigated by precise pulse recalibration and estimation of the spin-spin and spin-lattice relaxation times. The discussion here suggests that the large observed discrepancy may instead stem from structural factors, such as mosaicity and twin boundaries, or systematic errors in the spectral analysis, both of which warrant careful consideration.

\begin{table}[h]
\centering
\resizebox{\columnwidth}{!}{
\begin{tabular}{c c c} 
 \hline
& Suetsugu \textit{et al.}~\cite{Suetsugu2026} & Feng \textit{et al.}~\cite{Feng2023} \\ [0.5ex] 
 \hline\hline
$ \delta\nu_q $  $T\sim$ 100 K &  60 kHz  &  60 kHz \\
  \hline
 $ \delta\nu_q$  $T\sim$  180 K &  33 kHz &  20 kHz \\
  \hline
  Inherent $ \delta\nu_q$ ($\chi_0$) $T\sim \infty$  &  24 kHz &  14 kHz\\
 \hline
\end{tabular}}
\vspace{0.2cm}
\caption{\textbf{Comparison of $^{121}$Sb2 linewidths of the first NQR transition.} The baseline linewidth, which reflects intrinsic crystalline quality, is evaluated at temperatures well above $T_{\rm CDW}$. The $\sim 10$~kHz difference indicates a narrower intrinsic linewidth in the samples studied by Feng \textit{et al.}~\cite{Feng2023}, suggesting  the samples in a \cite{Feng2023} exhibited a higher degree of structural homogeneity.
} 
\label{table:1}
\vspace{-1cm}
\end{table}

In addition, the assertion of superior sample quality based on a sharper $^{121}$Sb2 NQR peak merits a nuanced examination~\cite{Suetsugu2026}. Just above the CDW transition, the reported linewidth of $\sim 60$~kHz in Ref.~\cite{Suetsugu2026} is identical to the value obtained by Feng \textit{et al.}~\cite{Feng2023}. To evaluate intrinsic sample quality, one must examine the baseline linewidth at high temperatures, far from $T_{\rm CDW}$, where pre-transitional fluctuations do not significantly distort the local EFG. 

In Tab.~\ref{table:1} we compare the reported $^{121}$Sb2 linewidth for the first NQR transition from~\cite{Feng2023, Suetsugu2026}. We also estimate the infinite-temperature linewidth, which directly measures inherent broadening from crystalline imperfections, via a CW-type fit:
\begin{equation}
\label{eq:CW}
\delta\nu(T) = \frac{C}{T-\theta_\text{CW}} + \chi_0 \, .
\end{equation}
This analysis reveals that the infinite-temperature broadening ($\chi_0$) in Suetsugu \textit{et al.} is substantially larger than that observed in the sample of Feng \textit{et al.}~\cite{Feng2023}. Additionally, the infinite-temperature broadening reported by Nikolov \textit{et al.} for the second NQR transition of $^{121}$Sb2 (spin-5/2) is consistent with the findings of Feng \textit{et al.} once the appropriate quantum mechanical scaling factor is applied~\cite{Nikolov2026}. 

This supports the magnitude of the simulated orientational disorder for Suetsugu \textit{et al.}'s sample with a fitted FWHM of $1.1^\circ$ at 160~K and $1.6^\circ$ at 100~K, reflecting respectively the baseline structural inhomogeneity and additional local structural distortions associated with pre-transitional CDW correlations. Although a baseline mosaicity of~$\lesssim 1.1^\circ$ indicates a crystal with satisfactory quality, it establishes a finite degree of structural inhomogeneity. Therefore,  when viewed alongside earlier measurements~\cite{Feng2023, Nikolov2026} that observed no anomalous features above the CDW transition, our analysis indicates that structural imperfections provide a robust and sufficient explanation for the Zp-NQR spectra reported by Suetsugu \textit{et al.}~\cite{Suetsugu2026}. 

\section{Hyperfine field model discussion}
\label{app:hyp_field}

Suetsugu et al.~\cite{Suetsugu2026} use a model of two hyperfine fields at the Sb2 site of $\Delta H = +h_{\rm loc}$ and $-\frac{1}{3}h_{\rm loc}$ in a $1\!:\!3$ ratio. Thus, they model the total spectral intensity using
\begin{equation}
\label{eq:}
I_{\mathrm{total}}(f,~\Delta)
=
\frac{A}{3}\,V(f+\Delta)
+
A\,V\!\left(f-\frac{\Delta}{3}\right),
\end{equation}
where $A$ is the normalization factor, $V(f)$ is the Voigt function representing the intrinsic lineshape of Sb2 and
\begin{equation}
\Delta = \gamma \mu_0 h_{\rm loc}
\end{equation}
is the frequency shift due to the local fields $h_{\mathrm{loc}}$.

In the Supplementary Information of~\cite{Suetsugu2026}, Suetsugu \textit{et al.} attribute the observed spectral broadening of the two magnetically inequivalent sites to thermal fluctuations of the local fields. However, constraints applied in the fitting routine are  not discussed in the main text or justified physically~\cite{Suetsugu2026}. Replicating the fitting curves provided in Ref.~\cite{Suetsugu2026} verifies that the model of two Voigt profiles $V(f)$ enforces a $3\!:\!1$ ratio for the intensity but a $1\!:\!1$ ratio for the linewidth between the two magnetic sites\footnote{See GitHub repository: \url{https://github.com/ilijanikolov/zpnqr_simulation}}. This equal-broadening constraint is surprising. If the linewidths arise from proportional distributions of the corresponding local fields, the widths would be expected to scale with the respective field magnitudes; an equal-linewidth constraint therefore imposes an additional assumption on the model. In fact, an equal-linewidth constraint is more naturally compatible with a common structural broadening mechanism than with independent distributions of inequivalent local magnetic fields.

\begin{figure}[t] 
\includegraphics[width=\linewidth]{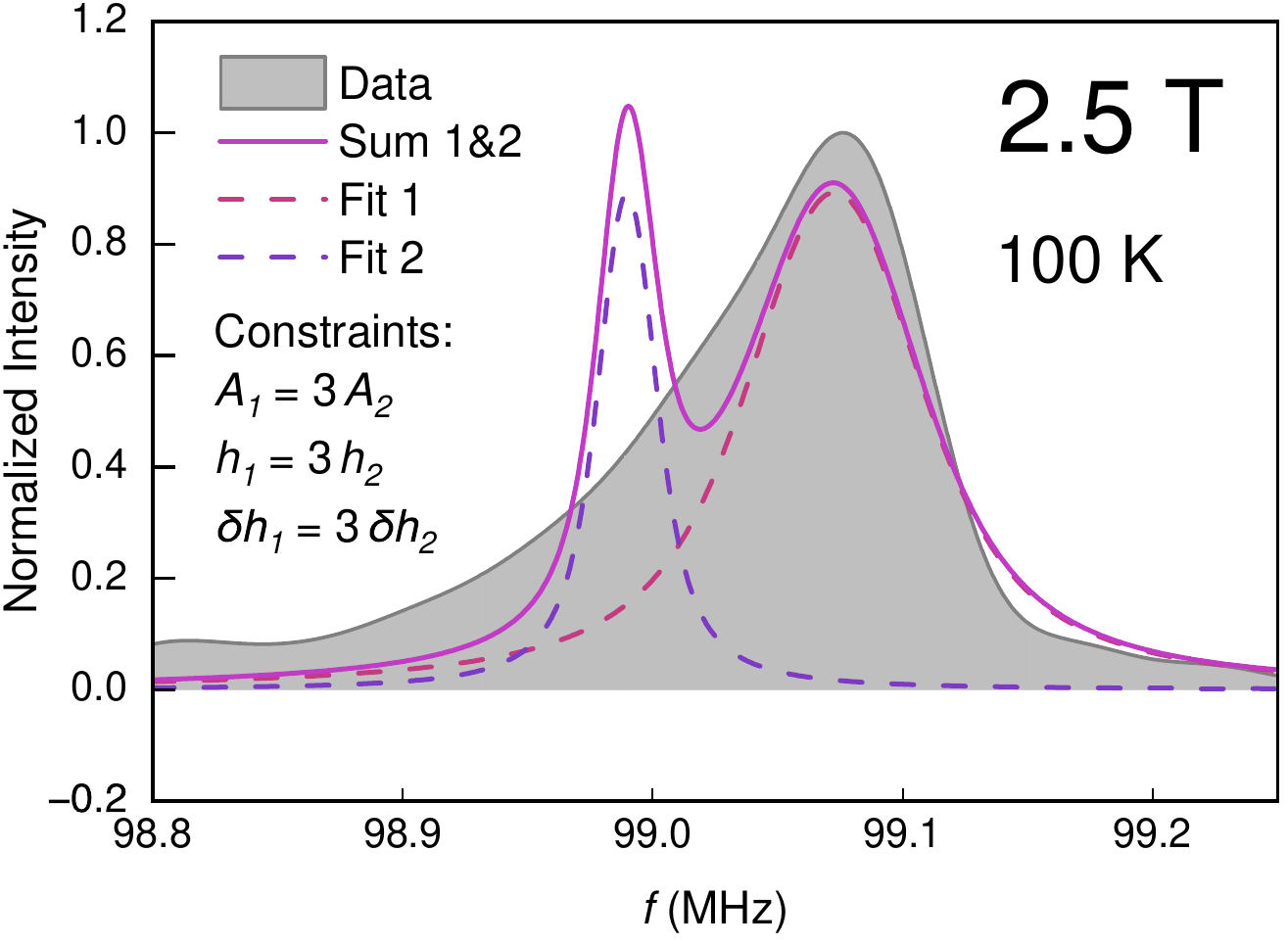}
\vspace{-.5cm}
\caption{\textbf{Broadening scaling in the loop-current model.} 
When the $3\!:\!1$ ratio for the two magnetically-inequivalent sites is applied to the intensity and linewidth, the simulation predicts a distinct spectral splitting (dashed lines), absent in the experimental data. The frequency constraint between the two peaks is given by $\Delta = \gamma\mu_0h_{\rm loc}$, where~$h_{\rm loc} \approx 65$~Oe as reported in~\cite{Suetsugu2026}.}
\label{fig:supp_constraints}
\vspace{-0.4cm}
\end{figure}

Upon relaxing the restrictive equal-broadening constraint but keeping the intensity ratio fixed to~$3\!:\!1$, fits to the experimental data demonstrate that the hyperfine field extrapolates to zero within uncertainty at zero applied field~(Fig.~\ref{fig:Fig_field}B). When the $3\!:\!1$ broadening and intensity constraints are applied, the fit produces a distinct, well-resolved secondary peak, not observed in the experimental data~(Fig.~\ref{fig:supp_constraints}). This suggests that the equal-broadening constraint mathematically leads the fit toward larger peak separations in order to reproduce the observed asymmetry, leading to an apparent finite $h_{\rm loc}$. 

\end{appendices}
\bibliography{ref}

\end{document}